\documentclass[conference]{IEEEtran}
\ifCLASSINFOpdf
\else
\fi

\usepackage[dvipsnames]{xcolor}
\usepackage{amsmath}
\usepackage{algorithm}
\usepackage[noend]{algpseudocode}
\usepackage{multicol}
\usepackage{pgfplots}
\usepackage{url}
\usepackage{tabularx, booktabs}

\newcommand{\eone}{\includegraphics[height=3ex]{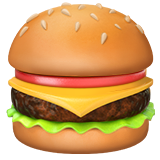}}
\newcommand{\etwo}{\includegraphics[height=3ex]{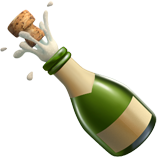}}
\newcommand{\ethree}{\includegraphics[height=3ex]{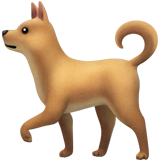}}
\newcommand{\efour}{\includegraphics[height=3ex]{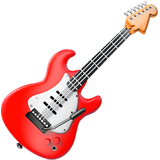}}
\newcommand{\efive}{\includegraphics[height=3ex]{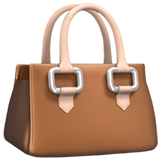}}
\newcommand{\esix}{\includegraphics[height=3ex]{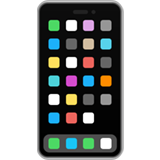}}
\newcommand{\eseven}{\includegraphics[height=3ex]{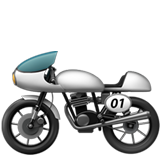}}
\newcommand{\eeight}{\includegraphics[height=3ex]{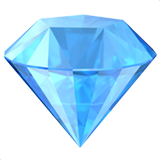}}
\newcommand{\enine}{\includegraphics[height=3ex]{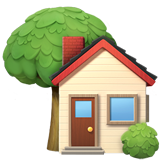}}
\newcommand{\eten}{\includegraphics[height=3ex]{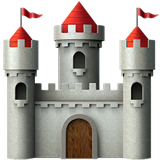}}

\newcommand{\tikzcircle}[2][red,fill=red]{\tikz[baseline=-0.5ex]\draw[#1,radius=#2] (0,0) circle ;}%
\pgfplotsset{compat=1.17}

\begin{document}
%
\title{Trust-ya: design of a multiplayer game \\ for the study of small group processes}

\author{\IEEEauthorblockN{Jerry Huang, Joshua Jung, Neil Budnarain, Benn McGregor and Jesse Hoey}
\IEEEauthorblockA{David R. Cheriton School of Computer Science\\
University of Waterloo\\
Waterloo, Ontario, Canada\\
Email: \{j354huan,j35jung,bbudnara,bamcgreg,jhoey\}@uwaterloo.ca}
}


%


\IEEEoverridecommandlockouts
\IEEEpubid{\makebox[\columnwidth]{978-1-6654-3886-5/21/\$31.00~\copyright2021 IEEE \hfill} \hspace{\columnsep}\makebox[\columnwidth]{ }}

\maketitle

\begin{abstract}
This paper presents the design of a cooperative multi-player betting game, {\em Trust-ya}, as a model of some elements of status processes in human groups. The game is designed to elicit status-driven leader-follower behaviours as a means to observe and influence social hierarchy. It involves a ``Bach/Stravinsky" game of deference in a group, in which people on each turn can either invest with another player or hope someone invests with them. Players who receive investment capital are able to gamble for payoffs from a central pool which then can be shared back with those who invested (but a portion of it may be kept, including all of it). The bigger gambles (people with more investors) get bigger payoffs. Thus, there is a natural tendency  for players to coalesce as investors around a ``leader" who gambles, but who also shares sufficiently from their winnings to keep the investors ``hanging on." The ``leader" will want to keep as much as possible for themselves, however. The game is played anonymously, but a set of ``status symbols" can be purchased which have no value in the game itself, but can serve as a ``cheap talk" communication device with other players. This paper introduces the game, relates it to status theory in social psychology, and shows some simple simulated and human experiments that demonstrate how the game can be used to study status processes and dynamics in human groups.
\end{abstract}


%
\IEEEpeerreviewmaketitle

\section{Introduction}

In games research, it is often desirable to design a mechanism such that agents acting selfishly will produce results with beneficial properties, like the maximization of overall welfare, or the equality with which wealth is distributed. However, in the real world, we may be forced to participate in `games' that tend toward undesirable outcomes, either due to their design, or the irrationality of their human players. It is therefore important to study games that allow players to act against the common good, and to understand how the addition of new agents can shift their behaviour.

Such games can often be cast as social dilemmas, such as the `Bach or Stravinsky' (BoS) problem.\footnote{In other literature, this problem may sometimes be referred to as the `battle of the sexes' or the `shopping/football problem'~\cite{Myerson91}.} In its basic form, BoS is a $2 \times 2$ matrix game that may be summarized as follows:

\begin{itemize}
    \item[] Two friends are interested in going to the symphony. One prefers Bach, while the other prefers Stravinsky, but it is most important to both that they go together. 
\end{itemize}

\noindent
The dilemma is clear, since both agents cannot have their preferred option while still being together, and one must settle for the other's preferred option in order to maximize utility. 
In this paper, we present a new game, called Trust-ya, that generalizes BoS to $N$ players and incorporates elements that expand the space of strategies, such as limited communication between players through the use of arbitrary symbols (emoji). The objective in the design of this game is to provide a platform for small-group research that captures the notions of status and durable inequality~\cite{Tilly1998,Burt1992,Ridgeway2019}, while remaining constrained enough to be approachable from a modeling perspective. According to Ridgeway~\cite{Ridgeway2019}, status is the esteem and influence granted by others, distinct from direct power or wealth. It is a kind of inequality that imposes a hierarchy on individuals, as well as groups within a society. In Trust-ya, emojis serve as status symbols, and the patterns by which they are accumulated allow us to observe how a social hierarchy can be formed and manipulated.

We describe the game, introduce simulated agents that map out the space of baseline gameplay, and give illustrative results from our (human) design team playing the game. Even in these simple demonstrations, we see the emergence of interesting and expected status behaviours. We conclude with a discussion of the social psychological background to the game, and present a mapping between game and theoretical elements. 

Ultimately, it may be possible to analyze group behaviours {\em a posteriori} in order to evaluate why a group failed to find a collaborative equilibrium, or to suggest mechanisms to achieve an equilibrium from an existing non-collaborative state. For example, such analysis may suggest that 
a particular kind of artificial agent be inserted in a way that induces an increase in efficiency. However, further research and analysis is needed on the dynamics, effects and ethics of such interventions. 

\section{Methods}

Trust-ya is a game in which 3 or more players aim to maximize their personal stash of coins drawn from a large central pile. Although they are in competition for these coins, players must cooperate to draw as many as possible from the central pile before the game ends. At each game step, a player may choose to either take $C_{take}$ coins from the central pile for themselves, or give $C_{give}$ coins from the central pile to any other player, where $C_{take}$ and $C_{give}$ are quantities that may be varied by the experimenter. Any player who receives coins (optionally, above some minimum threshold) may ``invest'' those coins and obtain a payout as a function of the invested amount. This function is scaled such that one player investing $x$ coins can expect a greater average payout than $n$ players each investing $x/n$ coins. The investing player may then decide how to distribute this payout among those players who contributed. This process is repeated until a stopping condition is met (most commonly, when the central pile is empty).

\begin{figure}
    \centering
    \includegraphics[width=
    \columnwidth]{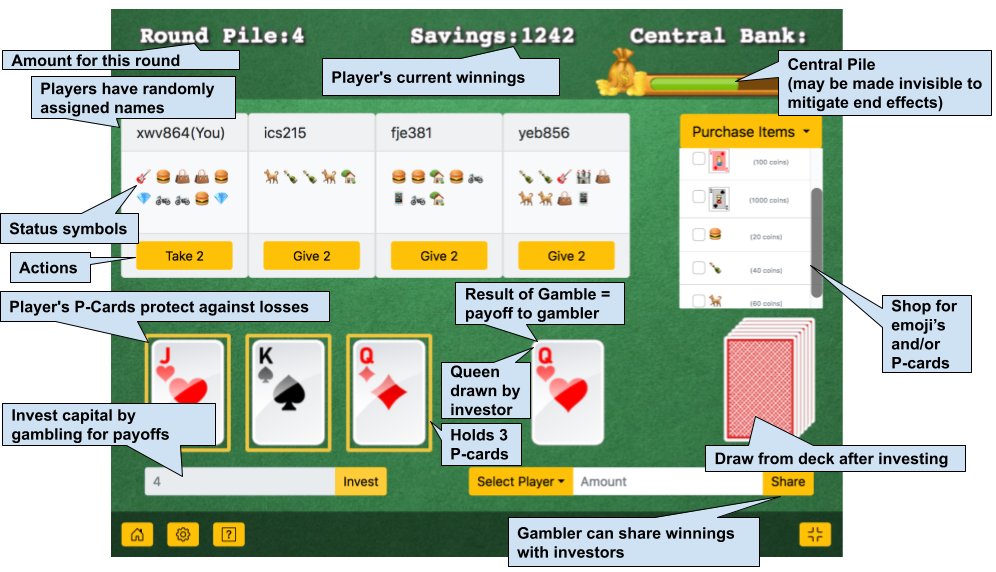}
    \vspace{-0.6cm}
    \caption{Annotated game interface at a central point in a game.}
    \label{fig:gameinterface}
    \vspace{-0.4cm}
\end{figure}

Between rounds, players may spend some of their accumulated coins to purchase ``P-cards'' (which provide investment benefits, as described in Section \ref{sec:details}), or ``status symbols,'' which have no intrinsic value, but serve as the only medium of communication between players. These symbols are currently displayed as emoji next to a player's name (from the set  (\begin{footnotesize}\eone\etwo\ethree\efour... \efive\esix\eseven\eeight\enine\eten\end{footnotesize}) but could be represented in a number of ways. The game interface is shown in Figure~\ref{fig:gameinterface} as a snapshot at about the mid-way point of a game.

\subsection{Game details \label{sec:details}}

For a group of $N$ players, the central pile initially contains $C_{ppp} \cdot N$ coins, where $C_{ppp}$ is the per-player pot size.
Each player begins with an empty savings pile, and 
can see the names and status symbols of all players. Other information, like each player's accumulated wealth, remains hidden. After all players have chosen to give or take coins in a round, they are each able to see who gave coins to them, but will only see the investment results of any player to whom they gave coins.



When a player makes an investment, a card is drawn from a standard 52-card deck. For non-face cards, the amount received increases with the size of the investment and the value of the card drawn, according to the function:

\vspace{-0.2cm}
\begin{equation} \label{eq:payout}
    \text{pay}(i) = \text{round}[\max(2^{i/2}-1, 0) \cdot v \cdot \alpha]
\end{equation}
\vspace{-0.5cm}

\noindent
where
$i$ is the number of coins invested, $v$ is the value of the card and $\alpha=50/N$ is a scaling factor which is decreasing with the number of players in the game. The scaling factor is used in order to allow the game to proceed much faster in the case of very small games, as in these cases the network dynamics tends to stabilize quickly, and the progression of the game becomes unnecessarily slow without it. This function is such that
winnings increase exponentially with the investment, which incentivizes players to attempt to form as large a group as possible. 
Further, the stochastic nature of drawing cards mimics uncertainty in real-life investments, and is necessary to introduce variance in payouts. Without such variance, it is too easy for players to make unthinking, repetitive decisions.


If a face card is drawn, a payout is only received if the investor possesses the corresponding ``P-card'' (protection card), which can be purchased between rounds with saved coins. There are three such cards, one for each face card, where the cost of the Jack P-card is lower than the Queen, which is lower than the King. If the investor has a P-card corresponding to the drawn card, then they receive a payout according to Equation \ref{eq:payout}, where $v$ may be set by the experimenter, but should generally follow the convention $v_{jack} \leq v_{queen} \leq v_{king}$.


If the player draws a face card while not in possession of the corresponding P-card, they get no payout and additionally lose a number of saved coins determined by the function:

\vspace{-0.2cm}
\begin{equation} \label{eq:loss}
    \text{loss}(r,i) = \text{round}\bigg[(s + r - i) \cdot \frac{i}{r}\cdot \frac{v}{100}\bigg]
\end{equation}
\vspace{-0.4cm}

\noindent
where $r$ is the number of coins that were available for the player to invest that round (distinct from $i$, the number actually invested), and $s$ is the size of the player's savings pile.
All coins lost are returned to the central pile. 
By this formula, loss is determined by the fraction of coins bet in a round, rather than the absolute number, which avoids any incentive to attract fewer investors to reduce risk.


Once payouts have been determined for all investors, each must decide how to distribute the coins gained among their backers (i.e. all players who gave them coins that round). Buttons are provided for quickly sharing the wealth evenly, or sharing nothing, and any other split may be entered manually. At this stage, P-cards may be lost by players with few backers.
P-cards are lost due to a lack of influence, based directly on the number of supporters a player has.  




The game ends either when the central pile has been depleted, or sometime after a predetermined number of rounds has passed. In order to minimize the impact of a fixed number of rounds on strategy, there is a chance of termination on every turn after the limit. For the remainder of this paper, we will assume a round limit of $50$, with a termination chance of $10\%$. 

\subsection{Measuring Utility and Equality} \label{sec:measurements}

We evaluate the final outcome of a game by the total utility achieved among all players, and the equality with which that utility is distributed. Total utility is given by the total number of coins gained from the central pile, while equality is measured by the Gini coefficient.\footnote{Gini is as a metric for measuring wealth or income inequality~\cite{gini1936measure}.} 
%
%
The Gini coefficient takes a minimum value of $0$ when wealth is distributed completely evenly, and a maximum value of $1-1/N$ when one player hoards all wealth, and the remaining $N-1$ players receive nothing. For a Trust-ya, we define a ``good'' outcome to be one where total utility is maximized, and the Gini is minimized.

\section{Results}
\subsection{Simulations}
We first ran a set of agent-based tests on the game to establish a baseline for different levels of earnings and equality, as shown in Figure~\ref{fig:botsims}.
These simulations used values $C_{ppp} = 10000$, $C_{give} = 2$, $C_{take} = 2$, $v_{jack} = 10$, $v_{queen} = 25$, $v_{king} = 50$.
\usetikzlibrary{calc}
\begin{figure}[tb]
    \centering
    \resizebox{0.85\columnwidth}{!}{
    \begin{tikzpicture}
        \begin{axis}[
            xmin=-0.3,xmax=1.1,
            ymin=-0.15,ymax=1.25,
            grid=both,
            grid style={line width=.1pt, draw=gray!10},
            major grid style={line width=.2pt,draw=gray!50},
            axis lines=middle,
            minor tick num=5,
            enlargelimits={abs=0},
            axis line style={latex-latex},
            ticklabel style={font=\tiny, fill=white},
            label style={font=\scriptsize},
            x label style={at={(axis description cs:0.5,0)},anchor=north},
            y label style={at={(axis description cs:0,.5)},rotate=90,anchor=south},
            xlabel={Gini Index},
            ylabel={Total Earnings (Fraction of Original Pile)}]
        ]
        
            \addplot[fill=green, fill opacity=0.3, draw opacity=0, area legend]
            coordinates {
                (0, 1)
                (0, 0.8)
                (0.2, 0.8)
                (0.2, 1)
            };
            \addplot[fill=red, fill opacity=0.3, draw opacity=0, area legend]
            coordinates {
                (0.5, 0)
                (1, 0)
                (1, 0.5)
                (0.5, 0.5)
            } -| (current plot begin);
            
            \node[pin={[pin distance=0.2cm]135:{\scriptsize{Bad Sharing}}}, circle, fill,color=blue,minimum size=0.1cm,inner  sep=0.25pt] at (axis cs:0.9,0.01) {};
            \node[pin={[pin distance=0.2cm]135:{\scriptsize{Taking}}}, circle,fill,color=blue,minimum size=0.1cm,inner sep=0.25pt] at (axis cs:0,0.01) {};
            \node[pin={[pin distance=0.2cm]90:{\scriptsize{Lucky Sharing}}}, circle,fill,color=red,minimum size=0.1cm,inner sep=0.25pt] at (axis cs:0.01854,1) {};
                
            \node[pin={[pin distance=0.2cm]315:{\scriptsize{Smart}}}, circle,fill,color=blue,minimum size=0.1cm,inner  sep=0.25pt] at (axis cs:0.59419788, 0.45908655) {};
            \node[pin={[pin distance=0.2cm]225:{\scriptsize{Smart/Random}}}, circle,fill,color=blue,minimum size=0.1cm,inner sep=0.25pt] at (axis cs:0.552476476931, 0.45639023) {};
            \node[pin={[pin distance=0.2cm]315:{\scriptsize{Smarter}}}, circle,fill,color=blue,minimum size=0.1cm,inner  sep=0.25pt] at (axis cs:0.33077298915, 0.81970756) {};
            \node[pin={[pin distance=0.2cm]0:{\scriptsize{Smarter/Random}}},circle,fill,color=blue,minimum size=0.1cm,inner sep=0.25pt] at (axis cs:0.3405312, 0.85179552) {};

            \node[pin={[pin distance=0.2cm]180:{\scriptsize{Status}}},circle,fill,color=blue,minimum size=0.1cm,inner  sep=0.1pt] at (axis cs:0.33184137,0.21743136) {};
            \node[pin={[pin distance=0.2cm]0:{\scriptsize{Smart Status}}},circle,fill,opacity=0.6,color=blue,minimum size=0.1cm,inner  sep=0.1pt] at (axis cs:0.04937,1) {};
            

        
        \end{axis}
    \end{tikzpicture}
    
    }
    \vspace{-0.3cm}
    \caption{Equality (Gini) and Wealth (Total Earnings) for simulations with a single agent type (see text). Green area indicates best outcomes on both dimensions (to within 20\% of the maximum possible earnings, and $G \leq 0.2$).
    Red area indicates worst outcomes (50\% of the maximum, and $G \geq 0.5$).  The optimal bot~\tikzcircle{2pt} for homogeneous populations is shown.
    }
    \label{fig:botsims}
    \vspace{-0.5cm}
\end{figure}
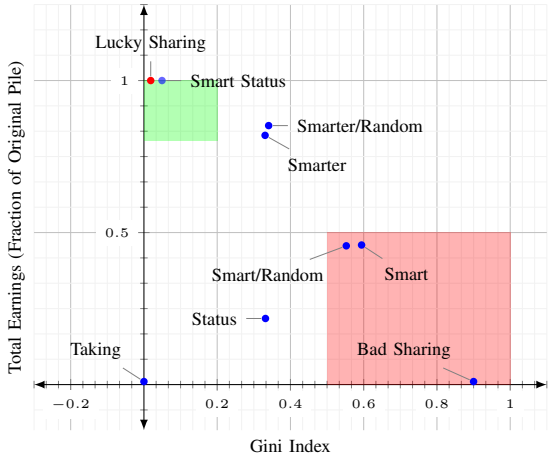

While ({\em Bad Sharing}, {\em Lucky Sharing}, and {\em Taking}) used an artificial coordination mechanism, the others did not. All agents take every round in {\em Bad Sharing}, leading to a very equitable, very inefficient payoff. {\em Bad Sharing}, in which all give to one randomly chosen, who never invests, shares or purchases, does badly on both counts. {\em Lucky Sharing} is the same but the lucky player invests everything, shares evenly, and buys P-cards. This is the best possible solution in terms of efficiency and equality. The non-coordinated bots with settings {\em Smart}, {\em Smart/Random}, {\em Smarter}, and {\em Smarter/Random} use a simple reinforcement learning strategy, giving coins to the bot with the best perceived rate of return based only on interactions with them. {\em Smart} bots buy P-cards when they can, {\em Smarter} bots only buy P-cards if they estimate they will be able to keep them, while the versions with {\em Random} in them give their coins to a random other player 5\% of the time. While the randomness makes little difference, likely due to the limited length of the games, the reticence for buying P-cards helps with earnings, and reduces inequality. {\em Status} and {\em Smart Status} bots attempt to invest in the player with the most emojis/status symbols, and are also not coordinated. A {\em Status} bot will attempt to buy an emoji every round, while a {\em Smart Status} bot will only do so if it received coins from at least one other player in the last round. {\em Status} bots perform poorly, as a leader cannot be distinguished when all players have the same number of emoji, but the {\em Smart} version do not have this problem, and so achieve high total utility and low inequality.


These bots simulations are done with all bots playing the same strategy to establish these as baselines. More complex interactions may be simulated, leading to a myriad of combinations in which different strategies interact to produce different outcomes. It is precisely these more complex interactions that form the basis of the emergence of leaders in the human tests (see next Section).

\subsection{Human Pilot Tests}
\label{sec:humantests}
\begin{figure}
    \centering
\usetikzlibrary{shapes,arrows,automata,plotmarks,positioning} 
\begin{tikzpicture}[main/.style = {draw, circle}]

\begin{scope}[scale=0.49, transform shape]
\node[main,minimum size=1.75cm,align=center] (a7) {\colorbox{BurntOrange}{vkq948}};
\node[right= 0.5cm of a7,main,minimum size=1.75cm,align=center] (a1) {\colorbox{Goldenrod}{cxf410}};
\node[right= 0.5cm of a1,main,minimum size=1.75cm,align=center] (a3) {\colorbox{LimeGreen}{qrv658}};
\node[right= 0.5cm of a3,main,minimum size=1.75cm,align=center] (a4) {\colorbox{GreenYellow}{ydm791}};
\node[right= 0.5cm of a4,main,minimum size=1.75cm,align=center] (a5) {\colorbox{yellow}{zlr097}};
\node[right= 0.5cm of a5,main,minimum size=1.75cm,align=center] (a2) {\colorbox{ProcessBlue}{xiy009}};
\node[right= 0.5cm of a2,main,minimum size=2.25cm,align=center] (a8) {\colorbox{SeaGreen}{epv816}};

\node[above= 0.5cm of a1] (o1) {\colorbox{Goldenrod}{cxf410}};
\node[above= 0.25cm of a8] (o2) {\colorbox{ProcessBlue}{xiy009}};
\node[above= 0.5cm of a5] (o3) {\colorbox{LimeGreen}{qrv658}};
\node[above= 0.5cm of a4] (o4) {\colorbox{GreenYellow}{ydm791}};
\node[above= 0.5cm of a3] (o5) {\colorbox{yellow}{zlr097}};
\node[right= 0.1cm of o2] (o6) {\colorbox{Periwinkle}{wsp502}};
\node[above= 0.5cm of a7] (o7) {\colorbox{BurntOrange}{vkq948}};
\node[above= 0.5cm of a2] (o8) {\colorbox{SeaGreen}{epv816}};

\draw[->] (o1) to (a1);
\draw[->] (o2) to (a8);
\draw[->] (o3) to (a5);
\draw[->] (o4) to (a4);
\draw[->] (o5) to (a3);
\draw[->] (o6) to (a8);
\draw[->] (o7) to (a7);
\draw[->] (o8) to (a2);

\node[below= 0.6cm of a7,main,minimum size=1.75cm,align=center] (b7) {\colorbox{BurntOrange}{vkq948}};
\node[right= 0.5cm of b7,main,minimum size=1.75cm,align=center] (b1) {\colorbox{Goldenrod}{cxf410}\\\eone};
\node[right= 0.5cm of b1,main,minimum size=1.75cm,align=center] (b4) {\colorbox{GreenYellow}{ydm791}};
\node[right= 0.5cm of b4,main,minimum size=1.75cm,align=center] (b5) {\colorbox{yellow}{zlr097}};
\node[right= 0.5cm of b5,main,minimum size=1.75cm,align=center] (b2) {\colorbox{ProcessBlue}{xiy009}};
\node[right= 0.5cm of b2,main,minimum size=2.25cm,align=center] (b8) {\colorbox{SeaGreen}{epv816}};

\draw[->] (a1) to node[] {} (b1);
\draw[->] (a5) to node[] {} (b5);
\draw[->] (a4) to node[] {} (b4);
\draw[->] (a3) to node[sloped,above] {\colorbox{yellow}{zlr097}} (b1);
\draw[->] (a7) to node[] {} (b7);
\draw[->] (a2) to node[] {} (b2);
\draw[->] (a8) to node[] {} (b8);

\node[below= 1.4cm of b1,main,minimum size=2.25cm,align=center] (c1) {\colorbox{Goldenrod}{cxf410}\\\eone};
\node[right= 0.5cm of c1,main,minimum size=2.25cm,align=center] (c3) {\colorbox{LimeGreen}{qrv658}\\\eone\etwo};
\node[right= 0.5cm of c3,main,minimum size=1.75cm,align=center] (c5) {\colorbox{yellow}{zlr097}};
\node[right= 0.5cm of c5,main,minimum size=1.75cm,align=center] (c2) {\colorbox{ProcessBlue}{xiy009}\\\etwo};
\node[right= 0.5cm of c2,main,minimum size=2.25cm,align=center] (c8) {\colorbox{SeaGreen}{epv816}\\\eseven};

\draw[->] (b1) [text width=2.5cm,midway,above=3em ] to (c1);
\draw[->] (b1) to node[left,sloped,above] {\colorbox{yellow}{zlr097}} (c3);
\draw[->] (b5) [text width=2.5cm,midway,above=3em ] to (c5);
\draw[->] (b4) to node[sloped,above] {\colorbox{GreenYellow}{ydm791}} (c3);
\draw[->] (b7) to node[right,sloped,above] {\colorbox{BurntOrange}{vkq948}} (c1);
\draw[->] (b2) [text width=2.5cm,midway,above=3em ] to (c2);
\draw[->] (b8) [text width=2.5cm,midway,above=3em ] to (c8);

\node[below=0.25cm of c1] (dots1) {\textbf{\dots}};
\node[below=0.6cm of c2] (dots2) {\textbf{\dots}};

\node[below= 0.7cm of c1,main,minimum size=4.5cm,align=center] (f1) {\colorbox{Goldenrod}{cxf410} \\ 
investors: 6\\
savings: 19133\\
P-cards: J,Q,K\\      \eone\ethree\eone\ethree\eone\\\eeight\eone
\eeight\eone\\\ethree\eone\ethree\eone};
\node[below= 1.9cm of c2,main,minimum size=2.25cm,align=center] (f2) {\colorbox{ProcessBlue}{xiy009} \\
investors: 2\\
savings: 5622\\
P-cards:2xJ,3xQ,K \\
\\\etwo\eone\ethree};

\node[below= 0.3cm of f1,main,minimum size=2.25cm,align=center] (g1) {\colorbox{Goldenrod}{cxf410}\\
investors: 5\\
savings: 20847\\
P-cards: J,Q,K \\
\\\eone\ethree\eone\ethree\eone\\\eeight\eone\eeight\eone\\\ethree\eone\ethree\eone};
\node[right= 0.2cm of g1,main,minimum size=1.65cm,align=center] (g3) {\colorbox{LimeGreen}{qrv658} \\
investors: 1\\
savings: 9620\\
P-cards: \\
\\\eone\etwo\eone\\\eone\eone\etwo\\\eone\etwo\etwo\\\etwo\etwo\eone};
\node[right= 0.2cm of g3,main,minimum size=1.65cm,align=center] (g2) {\colorbox{ProcessBlue}{xiy009} \\
investors: 1 \\
savings: 5624\\
P-cards: 2xJ,3xQ\\
\\\etwo\eone\ethree};
\node[right= 0.2cm of g2,main,minimum size=1.65cm,align=center] (g8) {\colorbox{SeaGreen}{epv816} \\
investors: 1\\
savings: 714 \\
P-cards: Q \\
\\\eseven\eten\\\eeight\enine};

\draw[->] (f1) [text width=2.5cm,midway,above=3em ] to (g1);
\draw[->] (f1) to node[right,sloped,above] {\colorbox{BurntOrange}{vkq948}} (g3);
\draw[->] (f2) [text width=2.5cm,midway,above=3em ] to (g2);
\draw[->] (f2) to node[right,sloped,above] {\colorbox{SeaGreen}{epv816}} (g8);
\end{scope}
\end{tikzpicture}
\vspace{-0.7cm}
    \caption{\label{fig:humanpilot} Game play of an all-human game with eight players showing emoji and groups formed. Nineteen mid-game stages are omitted, in which there were no significant changes other than increased emoji purchases, and a slow migration of most investors towards {\tt cxf410}. Nodes show the number of investors in a player, and that player's savings pile, P-cards, and status symbols.}
\vspace{-0.5cm}
\end{figure}
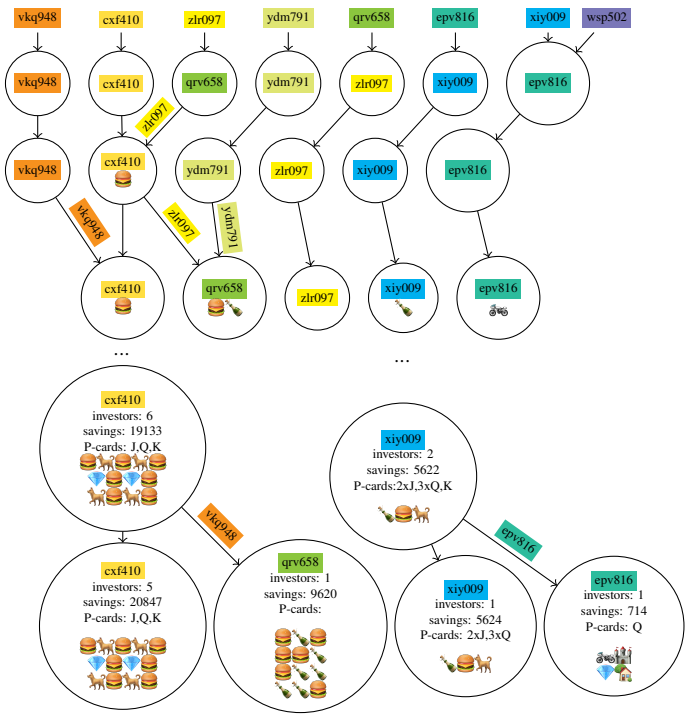

A team of researchers from the Computational Health Informatics Laboratory at the University of Waterloo were part of the overall research project, including the authors and 5 others including one postdoctoral fellow, three undergraduate students, and two graduate students, assisted in a number of design phases on a rolling basis. The number of participants in our design sessions varied from 4-14. During each session, we briefed all participants on any changes to the game since the last time. We then played a single game until the outcome of the game was clear or time ran out (approx. 1/2 hour). All participants were debriefed, and then we drew conclusions about next design steps. We engaged in about 6 of these sessions, and then ran two fully anonymized sessions of 1/2 hour each with 7 and 8 participants respectively. These games continued until the central pile was empty, and everyone's savings pile points were entered into a raffle for a pair of \$20 gift cards.\footnote{These games were played using a simpler interface than that shown in Figure~\ref{fig:gameinterface} (e.g. the images of playing cards and colors were missing), but the elements of the game were the same. To maintain a J,Q,K P-card a player needed $1$, $N/4$ and $N/2$ supporters, respectively.} The game play for one of these tests is shown in Figure \ref{fig:humanpilot}. We see the players initially start with a few small groups, but one player ({\tt cxf410}) attracted multiple others and was able to protect himself with many P-cards. Gameplay in the other games was similar, with one or two "leaders" emerging quickly, who sometimes engaged in a "status battle" as they try to out-buy the other in emojis. These battles were usually resolved when one or the other hit a King or Queen. The distribution of wealth in the human games typically ended with one player with 1/4 of the wealth, about half the players with half the wealth, and the remaining half with little. 

\section{Discussion}
A key element in group formation is the emergence of {\em leaders}, and therefore, also {\em followers}.
Leaders, or "the elite" usually obtain more resources and wield the power to shape social structure,
and thereby solidify their position as leaders~\cite{Piketti2014}. Trust-ya encodes some of the processes that underlie this status formation process, in the hopes that it can be used as a simulation environment to better understand them.

We ground Trust-ya in the status processes succinctly formulated by Ridgeway~\cite{Ridgeway2019}, who discusses dominance hierarchies (e.g. leader-follower groups), and shows how they require perceptions of perceived competence, as well as status beliefs about this same competence of other group members. 
Ridgeway's theory in some sense sums up a long sociological investigation that is referred to broadly as {\em expectation states theory}, broadly based on the the idea of reality, and status in particular, as something both objective (related to the real world) and subjective (related to the social structure of the world for any agent)~\cite{BergerLuckmann1967}. In expectation states theory, the dynamics of status follows a reflexive feedback loop. First, diffuse status characteristics are assigned based on observable status cues, such as gender, race, or physical artifacts: {\em ``...people jockey for position in the micro-dynamics of status through objects they display"}~\cite[p.113]{Ridgeway2019}. Second, these diffuse status characteristics modify expectations for performance in ways that are self-fulfilling. Agents with initially assigned high status are able, by using their increased resources, to define the direction the group takes in ways that are most conducive to maintaining the status hierarchy so defined. This status may then be challenged or found lacking in evidence by the group. Alternatively, low-status actors may be able to rise in status should they display some assertiveness, but not too much as this will be seen as disingenuous or deceptive. Leaders need to display competence, but also fairness and interest in a shared group identity with their followers 
\cite{Ridgeway2019}\cite{Steffens2020}.

In Trust-ya, status is observable only through each user's displayed emoji (and their random name). However, the objective status is really defined by a player's P-cards, which are hidden from the others (but possibly revealed during payouts). 
P-cards naturally reinforce an existing hierarchy in the social structure by forcing lower status groups to realize the expectations of higher status groups who control resources. The lower status group is then compelled to accept the meaning of the social relationship as defined by the higher status group, acceptance of which involves a re-definition of their position as one of deference. Higher status groups obtain more resources and thereby can better control the environment to suit their abilities, improving their status. To replicate this in Trust-ya, players with more P-cards get more resources when investing. 

The correspondence continues if one considers that the display of status is something agreed upon by both low and high status actors. Although Trust-ya is currently built using a ranked set of status symbols, these symbols may be stripped of meaning by offering them at an equal price. Meaning can then emerge as each symbol is
eventually reified by the group through gameplay. Over the course of our test games, we observed this process occurring for the hamburger (cheapest emoji). One player, Bob, bought many hamburger emoji in the first game, and ended up winning the game. Thereafter, the hamburger emoji became associated with winning and with Bob. 
In a game that supported dynamic, open market pricing, we speculate that our players may have been willing to pay more for the hamburger as a symbol of Bob-ness and of winning-ness, but this conjecture has yet to be demonstrated.

There are ethical concerns with the use of data from a game such as this to infer social status and dynamics for purposes outside of research. These concerns are not addressed in this paper, but 
we stress that the investigation of such issues is part of our ongoing efforts and plays a critical role in the development of Trust-ya~\cite{StarkHoey2020}.

\section{Conclusion}
This paper reviews the design of a game for investigating status processes in small groups. Our objective was to create a game that would replicate some elements of status processes, including the establishment of leader-follower behaviours. To do so, the game made use of a standard social dilemma, in which one agent must defer to another to achieve jointly optimal rewards. Coupled with a game that places one player in a position of dominance or control (the sharing mechanism), this naturally lead to the expected status orders being established in groups of players. We investigated the emergence of such orders in simulated agents and in humans, and we discussed theoretical connections to social psychology. Our aim is to further develop the game and its automated agents, and to observe their influence on the behaviour of human players.




\section*{Acknowledgment}

We thank the CHIL researchers who assisted with design iterations, and gratefully acknowledge support from AGE-WELL NCE Inc., the Alzheimer’s Association, and the Natural Sciences and Engineering Research Council of Canada.



%

\bibliographystyle{IEEEtran}
\bibliography{references.bib}

\end{document}